\documentclass[a4paper,11pt]{article}
\usepackage{cite}
\usepackage[all]{xy}
\usepackage{amsmath,amsfonts,amssymb,amsthm,epsfig,amscd,comment,latexsym,psfrag}
\usepackage{CJK}
\usepackage{mathrsfs}
\usepackage{nicefrac,xspace,tikz}
\usepackage{arydshln}
\usepackage{stmaryrd}
\usepackage{graphicx,bm}
\usepackage{amscd}
\usepackage{cite}
\usetikzlibrary{arrows}

\makeatletter

\newcommand{\Rmnum}[1]{\expandafter\@slowromancap\romannumeral #1@}
\makeatother
\usepackage{graphicx,amssymb,amsmath,bm,latexsym}
\allowdisplaybreaks

\begin{document}
\begin{center}
{\Large\bf Superintegrability for ($\beta$-deformed) partition function hierarchies with $W$-representations}\vskip .2in
{\large Rui Wang$^{a,}$\footnote{wangrui@cumtb.edu.cn},
Fan Liu$^{b,}$\footnote{liufan-math@cnu.edu.cn},
Chun-Hong Zhang$^{c,}$\footnote{zhangchunhong@ncwu.edu.cn},
Wei-Zhong Zhao$^{b,}$\footnote{Corresponding author: zhaowz@cnu.edu.cn}} \vskip .2in
$^a${\em Department of Mathematics, China University of Mining and Technology,
Beijing 100083, China}\\
$^b${\em School of Mathematical Sciences, Capital Normal University,
Beijing 100048, China} \\
$^c${\em School of Mathematics and Statistics, North China University of Water Resources and Electric Power,
Zhengzhou 450046, Henan, China}\\

\begin{abstract}
We construct the ($\beta$-deformed) partition function hierarchies with $W$-representations.
Based on the $W$-representations, we analyze the superintegrability property and derive their character
expansions with respect to the Schur functions and Jack polynomials, respectively.
Some well known superintegrable matrix models such as the Gaussian hermitian one-matrix model
(in the external field), $N\times N$ complex matrix model, $\beta$-deformed Gaussian hermitian
and rectangular complex matrix models are contained in the constructed hierarchies.

\end{abstract}
\end{center}

{\small Keywords: Matrix Models, Superintegrability}

\section{Introduction}

Recently there has been increasing interest in the superintegrability for matrix models \cite{AMironov1705,MironovJHEP082018,Morozov1901,Cassia2020,Shakirov2009,
MironovEurPhys,2203.03869,LY,2112.11371,Cordova,2107.13381,Mironovsummary,Mironov2105,
Mironov2104,Wangr2022,Mishnyakov2022,2204.14074,2206.02045}.
The superintegrability means that for the character expansions of the matrix models,
the average of a properly chosen symmetric function is proportional to ratios of symmetric functions
on a proper locus, i.e., $< character > \sim character$.
A wide range of matrix models are known to be superintegrable, such as
the (deformed) Gaussian hermitian and complex matrix models \cite{AMironov1705,MironovJHEP082018,Morozov1901,Cassia2020},
(Hurwitz-)Kontsevich matrix models \cite{Shakirov2009,MironovEurPhys}, unitary matrix models \cite{2203.03869},
fermionic matrix models \cite{LY,2112.11371}, and even some non-Gaussian matrix models \cite{Cordova,2107.13381}.
The constraints for matrix models are useful to analyze the structures of matrix models.
For the Gaussian hermitian one-matrix model, its character expansion with respect to the Schur functions
can be derived recursively from a single $w$-constraint \cite{Mironov2105}.
There are the Virasoro constraints (with higher algebraic structures) for matrix models. They
can be applied to analyze the character expansions of the matrix models as well, such as
the Gaussian hermitian one-matrix, complex matrix and fermionic matrix models \cite{LY,Mironov2104}.

$W$-representations of matrix models give the dual expressions
for the partition functions through differentiation rather than integration \cite{Shakirov2009}.
More precisely, the partition functions are realized by acting on elementary
functions with exponents of the given $W$-operators.
For the Gaussian tensor model \cite{Itoyama2020} and (fermionic) rainbow tensor models \cite{LY,Kang2021},
they can still be expressed as the $W$-representations.
Recently it was shown that the superintegrability for ($\beta$-deformed) matrix models
can be analyzed from their $W$-representations \cite{Wangr2022,Mishnyakov2022}.
In this paper, we will construct the partition function hierarchies with $W$-representations and analyze
the superintegrability property.

\section{Partition function hierarchies with $W$-representations}
 The Hurwitz-Kontsevich matrix model is a deformation of the Kontsevich model, which can be used to
describe the Hurwitz numbers
and Hodge integrals over the moduli space of complex curves \cite{Shakirov2009,Goulden97,KHurwitz}.
It is the special case of the more general Hurwitz partition functions \cite{0904.4227,1012.0433,Alexandrov2014}.
The Hurwitz-Kontsevich model is generated by the exponent of the Hurwitz operator $W_0$ acting on the function
$e^{p_1/e^{tN}}$,
\begin{eqnarray}\label{HKPF}
Z_{0}\{p\}=e^{tW_0}\cdot e^{p_1/e^{tN}}
=\sum_{\lambda}e^{tc_{\lambda}}S_\lambda\{p_k=e^{-tN}\delta_{k,1}\} S_{\lambda}\{p\},
\end{eqnarray}
where $t$ is a deformation
parameter, the Hurwitz operator $W_0$ is given by
\begin{eqnarray}
W_0=\frac{1}{2}\sum_{k,l=1}^{\infty}\big((k+l)p_{k}p_l
\frac{\partial}{\partial p_{k+l}}+klp_{k+l}\frac{\partial}{\partial p_k}\frac{\partial}{\partial p_l}\big)
+N\sum_{k=1}^{\infty}kp_{k}\frac{\partial}{\partial p_{k}},
\end{eqnarray}
and $c_{\lambda}=\sum_{(i,j)\in \lambda}(N-i+j)$.

The partition function  (\ref{HKPF}) possesses the matrix model representation \cite{Shakirov2009}
\begin{eqnarray}
Z_{0}\{p\}=\int_{N\times N} \sqrt{{\rm det}\left(\frac{{\rm sinh}(\frac{\phi\otimes I-I\otimes\phi}{2})}
{\frac{\phi\otimes I-I\otimes\phi}{2}} \right)}d\phi e^{-\frac{1}{2t}{\rm Tr}\phi^2
-\frac{N}{2}{\rm Tr}\phi-\frac{1}{6}tN^3+\frac{1}{24}tN+{\rm Tr}(e^{\phi}\psi)},
\end{eqnarray}
where  $\psi$ is an $N\times N$ matrix and the time variables $p_k={\rm Tr}\psi^k$.

Let us define the operator
\begin{equation}
E_{1}=[W_0, p_1]
=\sum_{n=1}^{\infty}np_{n+1}\frac{\partial}{\partial p_{n}}+ Np_1.
\end{equation}
Using the actions
\begin{equation}\label{W0action}
W_0 S_{\lambda}=c_{\lambda}S_{\lambda},
\end{equation}
and
\begin{equation}
p_1S_{\lambda}=\sum_{\lambda+\square}S_{\lambda+\square},
\end{equation}
we have
\begin{equation}
E_{1}S_{\lambda}=\sum_{\lambda+\square}(j_\square-i_\square+N) S_{\lambda+\square},
\end{equation}
 where $\lambda+\square$ are the Young diagrams obtained by adding one square  $\square=(i_\square, j_\square)$ to $\lambda$.

Let us set
\begin{equation}
W_{-1}=[W_0, E_{1}].
\end{equation}
The action of  $W_{-1}$ on the Schur functions is
\begin{eqnarray}
W_{-1}S_{\lambda}=\sum_{\lambda+\square}(j_\square-i_\square+N)^2 S_{\lambda+\square}.
\end{eqnarray}
In terms of $W_{-1}$ and $E_{1}$,  we introduce a series of operators
\begin{equation}
W_{-n}=\frac{1}{(n-1)!}\underbrace{[W_{-1}, [W_{-1}, \ldots [W_{-1} }_{n-1}, E_{1}]\ldots ]],\ \ n\geq 2.
\end{equation}

The actions of $W_{-n}$ with $n\geq 1$ on the Schur functions are given by
\begin{eqnarray}\label{actw-n}
W_{-n}S_{\lambda}=\sum_{\lambda+\square_1+\cdots+\square_n}\prod_{k=1}^{n}(j_{\square_k}-i_{\square_k}+N)^{\alpha}
A_{\lambda} ^{\lambda+\square_1+\cdots+\square_n}S_{\lambda+\square_1+\cdots+\square_n},
\end{eqnarray}
where we denote $\alpha=1+\delta_{n,1}$ in this paper for later convenience, $\lambda+\square_1+\cdots+\square_n$ are the
Young diagrams obtained by adding $n$ squares to $\lambda$, $A_{\lambda}^{\lambda+\square_1+\cdots+\square_n}$
are the coefficients in the actions
\begin{eqnarray}\label{multiply}
p_nS_{\lambda}=\sum_{\lambda+\square_1+\cdots+\square_n}
A_{\lambda}^{\lambda+\square_1+\cdots+\square_n}
S_{\lambda+\square_1+\cdots+\square_n}.
\end{eqnarray}
It is known that
$A_{\lambda}^{\lambda+\square_1+\cdots+\square_n}=\sum_{k=1}^n
(-1)^{k}a^{\lambda+\square_1+\cdots+\square_n}_{\lambda,(k,1^{n-k})}$,
 with the  Littlewood-Richardson coefficients  $a^{\lambda}_{\mu,\nu}$ defined by
$S_{\mu}S_{\nu}=\sum_{\lambda}a^{\lambda}_{\mu,\nu}S_{\lambda}$ \cite{Macdonaldbook}.
The actions (\ref{actw-n}) can be proved inductively by using the relations
$p_n=\frac{1}{n-1}[E_{1},p_{n-1}],\ \ n\geq 2$.

Let us introduce the partition function hierarchy with $W$-representations
\begin{equation}
Z_{-n}\{p\}=e^{{W}_{-n}/n}\cdot 1, \ \ n\geq 1.
\end{equation}

It is straightforward to calculate the powers of $W_{-n}$ acting on
$S_{\lambda}$ with $\lambda=\varnothing$, leading to the explicit results
\begin{eqnarray}\label{defactionwnm}
\frac{1}{n^mm!}{W}^m_{-n}\cdot 1=
  \sum_{\lambda\mapsto nm}
\left(\frac{S_{\lambda}\{p_k=N\}}
{S_{\lambda}\{p_k=\delta_{k,1}\}}\right)^{\alpha}
S_{\lambda}\{p_k=\delta_{k,n}\}
S_{\lambda},
\end{eqnarray}
where we have used the hook length formula
$\frac{S_{\lambda}\{p_k=N\}}
{S_{\lambda}\{p_k=\delta_{k,1}\}}=\prod_{(i,j)\in \lambda}
(j-i+N)$.

Then we have
\begin{equation}\label{zn}
Z_{-n}\{p\}=e^{{W}_{-n}/n}\cdot 1
=\sum_{\lambda}
\left(\frac{S_{\lambda}\{p_k=N\}}
{S_{\lambda}\{p_k=\delta_{k,1}\}}\right)^{\alpha}
S_{\lambda}\{p_k=\delta_{k,n}\}
S_{\lambda}\{p\}.
\end{equation}
We see that $Z_{-1}\{p\}$ and $Z_{-2}\{p\}$ are the
$N\times N$ complex matrix model \cite{AlexandrovJHEP2009,AMironov1705}
and Gaussian hermitian one-matrix model \cite{Shakirov2009,AMironov1705}, respectively,
\begin{eqnarray}\label{CPF}
Z_{-1}\{p\}&=&\frac{\int_{N\times N} d^2M e^{-{\rm Tr}MM^{\dagger}+\sum_{k=1}^{\infty}\frac{p_k}{k}{\rm Tr}(MM^{\dagger})^k}}
{\int_{N\times N} d^2M e^{-{\rm Tr}MM^{\dagger}}}\nonumber\\
&=&\sum_{\lambda}\frac{S_{\lambda}\{p_k=N\}^2}
{S_{\lambda}\{p_k=\delta_{k,1}\}}S_{\lambda}\{p\},\nonumber\\
Z_{-2}\{p\}&=&2^{-\frac{N}{2}}\pi^{-\frac{N^2}{2}}\int_{N\times N} dM e^{-\frac{1}{2}{\rm Tr}M^{2}+\sum_{k=1}^{\infty}
\frac{p_k}{k}{\rm Tr}M^k}\nonumber\\
&=&\sum_{\lambda}\frac{S_{\lambda}\{p_k=N\}}
{S_{\lambda}\{p_k=\delta_{k,1}\}}
S_{\lambda}\{p_k=\delta_{k,2}\}
S_{\lambda}\{p\}.
\end{eqnarray}

Note that for any partition function of the form \cite{Alexandrov2014}
\begin{equation}
\bar{Z}=\sum_{\lambda}\prod_{(i,j)\in{\lambda}}f(i-j)S_{\lambda}\{\bar p_k\}S_{\lambda}\{p_k\}
\end{equation}
with the arbitrary function $f$ and parameters $\bar p_k$, it is a $\tau$-function of the KP hierarchy.
It is clear that the partition function hierarchy (\ref{zn}) gives the $\tau$-functions of the KP hierarchy.

Similarly, we define the operator
\begin{equation}\label{E-1}
E_{-1}=[W_0, \frac{\partial}{\partial{p_{1}}}]
=-\sum_{n=1}^{\infty}(n+1)p_{n}\frac{\partial}{\partial p_{n+1}}- N\frac{\partial}{\partial p_1},
\end{equation}
and
\begin{equation}
E_{-1}S_{\lambda}=-\sum_{\lambda-\square}(j_\square-i_\square+N) S_{\lambda-\square},
\end{equation}
where $\lambda-\square$ are the Young diagrams obtained by removing one square
$\square=(i_\square, j_\square)$ from $\lambda$.

We set
\begin{eqnarray}\label{w1operator}
W_{1}=[W_0, E_{-1}].
\end{eqnarray}
There is the action
\begin{eqnarray}
W_{1}S_{\lambda}=\sum_{\lambda-\square}(j_\square-i_\square+N)^2 S_{\lambda-\square}.
\end{eqnarray}
 In terms of $W_{1}$ and $E_{-1}$, we introduce a series of operators
\begin{equation}\label{multibrawn}
W_{n}=\frac{(-1)^n}{(n-1)!}\underbrace{[W_{1}, [W_{1}, \ldots [W_{1} }_{n-1}, E_{-1}]\ldots ]],\ \ n\geq 2.
\end{equation}

The actions of $W_{n}$ on the Schur functions are given by
\begin{eqnarray}\label{actwn}
W_{n}S_{\lambda}=\sum_{\lambda-\square_1-\cdots-\square_n}
\prod_{k=1}^{n}
(j_{\square_k}-i_{\square_k}+N)^{\alpha}A_{\lambda} ^{\lambda-\square_1-\cdots-\square_n}
S_{\lambda-\square_1-\cdots-\square_n},\ \ n\geq 1,
\end{eqnarray}
where $\lambda-\square_1-\cdots-\square_n$ are the Young diagrams obtained
by removing $n$ squares from $\lambda$, $A_{\lambda}^{\lambda-\square_1-\cdots-\square_n}=\sum_{k=1}^{n}
(-1)^{k}a^{\lambda}_{\lambda-\square_1-\cdots-\square_n,(k,1^{n-k})}$ are the coefficients in the actions
\begin{eqnarray}\label{partial}
n\frac{\partial}{\partial p_n}S_{\lambda}=\sum_{\lambda-\square_1-\cdots-\square_n}
A_{\lambda}^{\lambda-\square_1-\cdots-\square_n}
S_{\lambda-\square_1-\cdots-\square_n}.
\end{eqnarray}
The actions (\ref{actwn}) can be proved inductively by using the relations
$\frac{\partial}{\partial p_n}=\frac{1}{n}[E_{-1},\frac{\partial}{\partial p_{n-1}}],\ \ n\geq 2$.
It is interesting to note that there are the same actions as (\ref{actwn}) for the operators
$W^-_n$ given by \cite{2206.02045}
\begin{equation}\label{wnmatrix}
W^-_n={\rm Tr}\frac{\partial^n}{\partial H^n}, \ \ n\geq 2,
\end{equation}
where $H$ is an $N\times N$ matrix.

Taking $p_k={\rm Tr}H^k$, we can rewrite the operators
(\ref{E-1}), (\ref{w1operator}) and $W_2$ in (\ref{multibrawn}) as
\begin{eqnarray}\label{w1matrix}
E_{-1}&=&-W^-_1=-{\rm Tr}\frac{\partial}{\partial H},\nonumber\\
W_1&=&\sum_{k,l=1}^{\infty}\big((k+l+1)p_{k}p_l
\frac{\partial}{\partial p_{k+l+1}}+klp_{k+l-1}\frac{\partial}{\partial p_k}\frac{\partial}{\partial p_l}\big)\nonumber\\
&&+2N\sum_{k=1}^{\infty}(k+1)p_{k}
\frac{\partial}{\partial p_{k+1}}
+N^2\frac{\partial}{\partial p_1}\nonumber\\
&=&{\rm Tr}(H^{T}\frac{\partial^2}{\partial H^2}),\nonumber\\
W_{2}&=&\sum_{k,l=1}^{\infty}\big((k+l+2)p_{k}p_l
\frac{\partial}{\partial p_{k+l+2}}+klp_{k+l-2}\frac{\partial}{\partial p_k}\frac{\partial}{\partial p_l}\big)\nonumber\\
&&+2N\sum_{k=1}^{\infty}(k+2)p_{k}
\frac{\partial}{\partial p_{k+2}}+2N^2\frac{\partial}{\partial p_{2}}+N\frac{\partial^2}{\partial p^2_{1}}\nonumber\\
&=&{\rm Tr}\frac{\partial^2}{\partial H^2},
\end{eqnarray}
where $H^{T}$ is the transpose of the matrix $H$.

Since (\ref{multibrawn}) can be expressed as
$W_{n}=\frac{1}{n-1}[W_{n-1},W_{1}], \  n\geq 3,$
it is clear that the operators $W_n$ (\ref{multibrawn}) and $W^-_n$ (\ref{wnmatrix}) are equivalent for $n\geq 2$.

Let us introduce the partition function hierarchy with $W$-representations
\begin{eqnarray}\label{zn2}
{  Z}_n\{g|p\}
=e^{W_{n}/n}e^{\sum_{k=1}^{\infty}
\frac{p_kg_k}{k}},\ \ n\geq 1.
\end{eqnarray}

Since there are the actions
\begin{eqnarray}
\frac{1}{n^m m!}W_{n}^m S_{\lambda}
=\sum_{\mu\mapsto |\lambda|-nm}
\left(\frac{S_{\lambda}\{p_k=N\}S_{\mu}\{p_k=\delta_{k,1}\}}
{S_{\mu}\{p_k=N\}S_{\lambda}\{p_k=\delta_{k,1}\}}\right)^{\alpha}
S_{\lambda/\mu}
\{p_k=\delta_{k,n}\}
S_{\mu},
\end{eqnarray}
where $S_{\lambda/\mu}$ are the skew Schur functions,
we obtain the character expansions for the partition function hierarchy (\ref{zn2})
\begin{eqnarray}\label{cezngp}
{{ Z}}_n\{g|p\}=\sum_{\lambda,\mu}
\left(\frac{S_{\lambda}\{p_k=N\}S_{\mu}\{p_k=\delta_{k,1}\}}
{S_{\mu}\{p_k=N\}S_{\lambda}\{p_k=\delta_{k,1}\}}\right)^{\alpha}
S_{\lambda/\mu}
\{p_k=\delta_{k,n}\}
S_{\lambda}\{g\}S_{\mu}\{p\}.
\end{eqnarray}
Here we have used the Cauchy formula
$e^{\sum_{k=1}^{\infty}\frac{1}{k}p_k\bar p_k}=\sum_{\lambda}S_{\lambda}\{p\}S_{\lambda}\{\bar p\}$.

When particularized to the $n=2$ case in (\ref{cezngp}), it gives the character
expansion \cite{Wangr2022} of Gaussian hermitian one-matrix model
in the external field \cite{Shakirov2009}
\begin{eqnarray}\label{HEPF}
Z_{2}\{g|p\}&=&\int dM_1 e^{-{\rm Tr}\frac{M_1^{2}}{2}+\sum_{k=1}^{\infty}\frac{g_k}{k}{\rm Tr}(M_1+M_2)^k}\nonumber\\
&=&\sum_{\lambda,\mu}\frac{S_{\lambda}\{p_k=N\}S_{\mu}\{p_k=\delta_{k,1}\}}
{S_{\mu}\{p_k=N\}S_{\lambda}\{p_k=\delta_{k,1}\}}
S_{\lambda/\mu}
\{p_k=\delta_{k,2}\}S_{\lambda}\{g\}S_{\mu}\{p\},
\end{eqnarray}
where $p_k={\rm Tr}M_2^k$.

\section{$\beta$-deformed partition function hierarchies with $W$-representations}
Let us extend the Hurwitz-Kontsevich matrix model (\ref{HKPF}) to the $\beta$-deformed case,
\begin{equation}
{\mathcal Z}_{0}\{p\}
=e^{t\mathcal{W}_0}\cdot e^{\beta p_1/e^{tN}},
\end{equation}
where
\begin{eqnarray}
\mathcal{W}_0&=&\frac{1}{2}\sum_{k,l=1}^{\infty}\big(\beta(k+l)p_{k}p_l
\frac{\partial}{\partial p_{k+l}}+klp_{k+l}\frac{\partial}{\partial p_k}\frac{\partial}{\partial p_l}\big)\nonumber\\
&&+\frac{1}{2}\sum_{k=1}^{\infty}\big((1-\beta)(k-1)+2\beta N\big)kp_{k}\frac{\partial}{\partial p_k}.
\end{eqnarray}

Taking $\bar p_k=e^{-tN}\delta_{k,1}$ in the Cauchy formula
\begin{equation}\label{cauchy}
e^{\beta\sum_{k=1}^{\infty}\frac{p_k\bar p_k}{k}}=\sum_{\lambda}\frac{1}
{\langle J_{\lambda}, J_{\lambda}\rangle_{\beta}}J_{\lambda}\{p_k\}
J_{\lambda}\{\bar p_k\},
\end{equation}
we have
\begin{equation}\label{dp1}
e^{\beta p_1/e^{tN}}=\sum_{\lambda}\frac{1}{\langle J_{\lambda}, J_{\lambda}\rangle_{\beta}}
J_{\lambda}\{p_k=e^{-tN}\delta_{k,1}\}J_{\lambda}\{p_k\}.
\end{equation}
Here
$\langle J_{\lambda}, J_{\lambda}\rangle_{\beta}=\frac{h_{\lambda}}{h^{'}_{\lambda}}$,
$h_{\lambda}=\prod_{(i,j)\in \lambda}(1+\lambda_i-j+\beta(\lambda_j^{'}-i))$ and
$h^{'}_{\lambda}=\prod_{(i,j)\in \lambda}(\lambda_i-j+\beta(\lambda_j^{'}-i+1))$ are the deformed hook length,
in which $\lambda^{'}=(\lambda_1^{'},\lambda_2^{'},\cdots)$ is the conjugate partition of $\lambda$.

Using the expansion (\ref{dp1}) and the action \cite{Lassalle2009}
\begin{eqnarray}\label{dW0action}
\mathcal{W}_0 J_{\lambda}=\bar c_{\lambda}J_{\lambda},
\end{eqnarray}
where $\bar c_{\lambda}=\sum_{(i,j)\in \lambda}c(i,j)$ and $c(i,j)=j-1+\beta (N-i+1)$, we then obtain
\begin{eqnarray}
\mathcal Z_{0}\{p\}
=\sum_{\lambda}\frac{e^{t\bar c_{\lambda}}}{\langle J_{\lambda}, J_{\lambda}\rangle_{\beta}}
J_\lambda\{p_k=e^{-tN}\delta_{k,1}\} J_{\lambda}\{p\}.
\end{eqnarray}

Let us define the operator
\begin{equation}
\bar E_{1}=[\mathcal{W}_0, p_1]
=\sum_{n=1}^{\infty}np_{n+1}\frac{\partial}{\partial p_{n}}+ \beta Np_1.
\end{equation}
We have
\begin{eqnarray}\label{dE1action}
\bar E_{1}J_{\lambda}=\sum_{\lambda+\square}c(i_\square,j_\square)
B_{\lambda}^{\lambda+\square} J_{\lambda+\square},
\end{eqnarray}
where $B_{\lambda}^{\lambda+\square}$ are the
coefficients in the action
\begin{eqnarray}\label{dactp1}
p_1J_{\lambda}=\sum_{\lambda+\square}B_{\lambda} ^{\lambda+\square}J_{\lambda+\square}.
\end{eqnarray}

Since $p_2=[\bar E_1, p_1]$, from the actions (\ref{dE1action}) and (\ref{dactp1}), we obtain
\begin{equation}\label{dactp2}
p_2J_{\lambda}=\sum_{\lambda+\square_1+\square_2}
B_{\lambda}^{\lambda+\square_1+\square_2} J_{\lambda+\square_1+\square_2},
\end{equation}
where $B_{\lambda}^{\lambda+\square_1+\square_2}=
(c(i_{\square_2},j_{\square_2})-
c(i_{\square_1},j_{\square_1}))B_{\lambda}^{\lambda+\square_1}
B_{\lambda+\square_1}^{\lambda+\square_1+\square_2}$.

For the operator
\begin{equation}
{\mathcal W}_{-1}=[\mathcal{W}_0, \bar E_{1}],
\end{equation}
 we have
\begin{equation}\label{dactw-1}
\mathcal W_{-1}J_{\lambda}=\sum_{\lambda+\square}c^2(i_\square,j_\square)
B_{\lambda}^{\lambda+\square} J_{\lambda+\square}.
\end{equation}
Note that the combination of $\mathcal W_{-1}$ and $\bar E_{1}$, i.e.,
${\mathcal W}_{-1}+(1-\beta)\bar E_{1}$, gives the $W$-representation of the
$\beta$-deformed $N\times N$ complex matrix model \cite{Cheny,Cassia2020}.

For the operator
\begin{equation}
{\mathcal W}_{-2}=[\mathcal{W}_{-1}, \bar E_{1}],
\end{equation}
 it gives the $W$-operator
in the $W$-representations of $\beta$-deformed Gaussian hermitian matrix model \cite{Morozov1901}.
By the actions (\ref{dE1action}) and (\ref{dactw-1}), we have
\begin{equation}\label{dactw-2}
\mathcal W_{-2}J_{\lambda}=\sum_{\lambda+\square_1+\square_2}
c(i_{\square_1},j_{\square_1}) c(i_{\square_2},j_{\square_2})
B_{\lambda}^{\lambda+\square_1+\square_2} J_{\lambda+\square_1+\square_2},
\end{equation}
where $B_{\lambda}^{\lambda+\square_1+\square_2}$ are the coefficients in (\ref{dactp2}).

Let us introduce a series of operators
\begin{equation}
\mathcal{W}_{-n}=\frac{1}{(n-1)!}\underbrace{[\mathcal{W}_{-1}, [\mathcal{W}_{-1}, \ldots [\mathcal{W}_{-1} }_{n-1},
\bar E_{1}]\ldots ]],\ \ n\geq 2.
\end{equation}
There are the actions
\begin{eqnarray}\label{dactw-n}
\mathcal W_{-n}J_\lambda=\sum_{\lambda+\square_1+\cdots+\square_n}
\prod_{k=1}^{n}c(i_{\square_k},j_{\square_k})^\alpha B_{\lambda} ^{\lambda+\square_1+\cdots+\square_n}
J_{\lambda+\square_1+\cdots+\square_n},\ \ n\geq 1,
\end{eqnarray}
where $B_{\lambda} ^{\lambda+\square_1+\cdots+\square_n}$ are the coefficients in
\begin{eqnarray}
p_nJ_{\lambda}=\sum_{\lambda+\square_1+\cdots+\square_n}B_{\lambda} ^{\lambda+\square_1+\cdots+\square_n}
J_{\lambda+\square_1+\cdots+\square_n}.
\end{eqnarray}
The actions (\ref{dactw-n}) can be proved inductively by using  the relations
$p_n=\frac{1}{n-1}[\bar E_{1},p_{n-1}],\ \ n\geq 2$.

Let us introduce the partition function hierarchy with $W$-representations
\begin{equation}
{\mathcal Z}_{-n}\{p\}= e^{\mathcal W_{-n}/n}\cdot 1, \ \ n\geq 1.
\end{equation}

It is straightforward to calculate the power of $\mathcal W_{-n}$ acting on
$J_{\lambda}$ with $\lambda=\varnothing$,  leading to the explicit result
\begin{eqnarray}
\mathcal W^m_{-n}\cdot 1=\sum_{\lambda\mapsto nm}\prod_{(i,j)\in \lambda} c(i,j)^{\alpha}b(\lambda)
J_{\lambda}, \ \ n\geq 1,
\end{eqnarray}
where
\begin{equation}
b(\lambda)=\frac{\langle p_n^m,J_{\lambda}\rangle_{\beta}}{\langle J_{\lambda}, J_{\lambda}\rangle_{\beta}}
=m!n^m\beta^{-m}\frac{J_{\lambda}\{p_k=\delta_{k,n}\}}{\langle J_{\lambda}, J_{\lambda}\rangle_{\beta}}.
\end{equation}
Using the hook length formula
\begin{equation}
\frac{J_{\lambda}\{p_k=N\}}{J_{\lambda}\{p_k=\delta_{k,1}\}}=\beta^{-|\lambda|}
\prod_{(i,j)\in \lambda}c(i,j),
\end{equation}
we further obtain
\begin{equation}\label{ceznp}
{\mathcal Z}_{-n}\{p\}
=\sum_{\lambda}\beta^{|\lambda|(n\alpha-1)/n}
\left(\frac{J_{\lambda}\{p_k=N\}}
{J_{\lambda}\{p_k=\delta_{k,1}\}}\right)^{\alpha}
\frac{J_{\lambda}\{p_k=\delta_{k,n}\}}{\langle J_{\lambda}, J_{\lambda}\rangle_{\beta}}
J_{\lambda}\{p\}.
\end{equation}

When particularized to the $n=1, 2$ cases in (\ref{ceznp}), it gives the $\beta$-deformed rectangular
complex (with $N_1=N_2$) and Gaussian hermitian matrix models \cite{Morozov1901}, respectively,
\begin{eqnarray}
{\mathcal Z}_{-1}\{p\}
&=&\sum_{\lambda}\beta^{|\lambda|}
\frac{J_{\lambda}\{p_k=N\}^2}
{J_{\lambda}\{p_k=\delta_{k,1}\}\langle J_{\lambda}, J_{\lambda}\rangle_{\beta}}
J_{\lambda}\{p\},\nonumber\\
{\mathcal Z}_{-2}\{p\}&=&\prod_{i=1}^N\int_{-\infty}^{+\infty} dz_i\Delta(z)^{2\beta}
e^{\sum_{i=1}^{N}\sum_{k=1}^{\infty}\beta\frac{ p_k}{k}z_i^k-
\frac{1}{2}\sum_{i=1}^Nz_i^2}\nonumber\\
&=&\sum_{\lambda}\beta^{|\lambda|/2}
\frac{J_{\lambda}\{p_k=N\}J_{\lambda}\{p_k=\delta_{k,2}\}}
{J_{\lambda}\{p_k=\delta_{k,1}\}
\langle J_{\lambda}, J_{\lambda}\rangle_{\beta}}
J_{\lambda}\{p\}.
\end{eqnarray}

Let us turn to construct the operator
\begin{equation}
\mathcal W_{1}=[\mathcal W_{0},{\bar E}_{-1}],
\end{equation}
where the operator $\bar E_{-1}$ is given by
\begin{equation}
\bar E_{-1}=\beta^{-1}[\mathcal W_0, \frac{\partial }{\partial p_1}]
=-\sum_{n=1}^{\infty}(n+1)p_{n}\frac{\partial}{\partial{p_{n+1}}}
-N\frac{\partial}{\partial{p_{1}}}.
\end{equation}
We have the action
\begin{eqnarray}
\mathcal W_{1}J_{\lambda}=\sum_{\lambda-\square}
c^{2}(i_\square,j_\square)B_\lambda^{\lambda-\square}J_{\lambda-\square},
\end{eqnarray}
where $B_\lambda^{\lambda-\square}$ are the coefficients in
\begin{eqnarray}\label{dpartial1}
\beta^{-1} \frac{\partial }{\partial p_1}J_{\lambda}=\sum_{\lambda-\square}
B_{\lambda}^{\lambda-\square}
J_{\lambda-\square}.
\end{eqnarray}

Let us introduce the operators
\begin{equation}
\mathcal{W}_{n}=\frac{(-1)^n}{(n-1)!}\underbrace{[\mathcal{W}_{1}, [\mathcal{W}_{1}, \ldots [\mathcal{W}_{1} }_{n-1},
\bar E_{-1}]\ldots ]],\ \ n\geq 2.
\end{equation}
We have the actions
\begin{eqnarray}
\mathcal W_{n}J_{\lambda}=\sum_{\lambda-\square_1-\cdots-\square_n}
\prod_{k=1}^{n} c(i_{\square_k},j_{\square_k})^\alpha
B_{\lambda}^{\lambda-\square_1-\cdots-\square_n}
J_{\lambda-\square_1-\cdots-\square_n}, \ \ n\geq 1,
\end{eqnarray}
where $B_{\lambda}^{\lambda-\square_1-\cdots-\square_n}$ are the coefficients in
\begin{eqnarray}\label{dpartial}
\beta^{-1} n\frac{\partial }{\partial p_n}J_{\lambda}=\sum_{\lambda-\square_1-\cdots-\square_n}
B_{\lambda}^{\lambda-\square_1-\cdots-\square_n}
J_{\lambda-\square_1-\cdots-\square_n}.
\end{eqnarray}

We introduce the partition function hierarchy with $W$-representations
\begin{eqnarray}\label{zngp}
{\mathcal Z}_n\{g|p\}=e^{\mathcal W_{n}/n}e^{\beta\sum_{k=1}^{\infty}
\frac{p_kg_k}{k}},\ \ n\geq 1.
\end{eqnarray}

Due to
\begin{eqnarray}\label{powwnjack}
\frac{1}{n^m m!}\mathcal W_{n}^m J_\lambda=\sum_{\mu\mapsto |\lambda|-nm}\beta^{m(n\alpha-1)}
\left(\frac{J_{\lambda}\{p_k=N\}J_{\mu}\{p_k=\delta_{k,1}\}}
{J_{\mu}\{p_k=N\}J_{\lambda}\{p_k=\delta_{k,1}\}}\right)^{\alpha}
\frac{J_{\lambda/\mu}\{p_k=\delta_{k,n}\}}
{\langle J_{\mu}, J_{\mu}\rangle_{\beta}}J_{\mu},
\end{eqnarray}
and the Cauchy formula (\ref{cauchy}),
there are the character expansions for the partition function hierarchy (\ref{zngp})
\begin{eqnarray}
{\mathcal Z}_n\{g|p\}=\sum_{\lambda,\mu}\beta^{|\lambda/\mu|(n\alpha-1)/n}
\left(\frac{J_{\lambda}\{p_k=N\}J_{\mu}\{p_k=\delta_{k,1}\}}
{J_{\mu}\{p_k=N\}J_{\lambda}\{p_k=\delta_{k,1}\}}\right)^{\alpha}
\frac{J_{\lambda/\mu}
\{p_k=\delta_{k,n}\}}
{\langle J_{\lambda}, J_{\lambda}\rangle_{\beta}\langle J_{\mu}, J_{\mu}\rangle_{\beta}}
J_{\lambda}\{g\}J_{\mu}\{p\}.
\end{eqnarray}

\section{Summary}
It was known that the Hurwitz-Kontsevich matrix model (\ref{HKPF}) can be expressed as the exponent of
the Hurwitz operator $W_0$ acting on the function $e^{p_1/e^{tN}}$. In terms of the Hurwitz operator $W_0$, $p_1$ and
$\frac{\partial }{\partial p_1}$, we have constructed the  partition function hierarchies with $W$-representations.
Based on the $W$-representations, we showed that these partition functions can be expressed as the character
expansions with respect to the Schur functions. It was noted that the character expansions of hierarchy (\ref{zn})
give the $\tau$-functions of the KP hierarchy, and the $N\times N$ complex matrix and Gaussian hermitian one-matrix
models are contained in the hierarchy (\ref{zn}). For the constructed partition function hierarchy (\ref{cezngp}),
it contains the Gaussian hermitian one-matrix model in the external field. We have also extended  the Hurwitz-Kontsevich
matrix model (\ref{HKPF}) to the $\beta$-deformed case. Similarly, the $\beta$-deformed partition function
hierarchies with $W$-representations were constructed and their character expansions with respect to the Jack polynomials
were presented as well. The $\beta$-deformed rectangular complex and Gaussian hermitian matrix models are contained in
the hierarchy (\ref{ceznp}). Searching for the matrix model representations of the partition functions in the hierarchies
presented in this paper would merit further investigations. Furthermore, it would be interesting to construct $q,t$-deformed
partition function hierarchies with $W$-representations and study their character expansions with respect to the
 Macdonald polynomials.

\section *{Acknowledgments}
We are grateful to A. Morozov and A. Mironov for their helpful comments.
This work is supported by the National Natural Science Foundation of China (Nos. 11875194 and 12105104)
and the Fundamental Research Funds for the Central Universities, China (No. 2022XJLX01).


\end{document}